\documentclass[aps,prl,twocolumn,amsfonts,amssymb,amsmath,showpacs]{revtex4}
\usepackage{graphicx}

\newcommand{\eg}[1]{{\it e.g.\/}\ifx#1.\else\expandafter#1\fi}
\newcommand{\ie}[1]{{\it i.e.\/}\ifx#1.\else\expandafter#1\fi}

\newcommand{\df}[2]{\frac{\partial #1}{\partial #2}}
\newcommand{\ddf}[2]{\frac{\partial^2 #1}{\partial #2^2}}
\newcommand{\e}{e}                 % nat log base
\newcommand{\eq}[1]{\eqref{#1}}
\newcommand{\eqtwo}[2]{(\ref{#1},\ref{#2})}
\renewcommand{\i}{i}               % imaginary unit
\newcommand{\Mx}[1]{\left[\begin{array}{cc}#1\end{array}\right]}
\newcommand{\mx}[1]{\mathbf{#1}}
\renewcommand{\Re}[1]{\mathop{\mathrm{Re}}\left(#1\right)}
\newcommand{\sech}{\mathop{\mathrm{sech}}}  % hyperbolic secant

\renewcommand{\a}{a}               % u-threshold parameter
\newcommand{\ale}{A}               % a-parameter in LE model
\newcommand{\A}{\mx{A}}            % linearization matrix
\newcommand{\ble}{B}               % b-parameter in LE model
\newcommand{\C}{C}                 % arb const
\newcommand{\cg}{c}                % group velocity
\newcommand{\cp}{c_{ph}}           % phase velocity
\newcommand{\decr}{\mu}            % spatial decrement of the front
\newcommand{\eps}{\varepsilon}     % v-slowness parameter
\newcommand{\evec}{\mx{v}}         % eigenvector
\newcommand{\evals}{\nu}           % spatial eigenvalue
\newcommand{\evalt}{\lambda}       % temporal eigenvalue
\newcommand{\f}{f}                 % u-kinetics
\newcommand{\freq}{\omega}         % ang frequency in the front
\newcommand{\g}{g}                 % v-kinetics
\newcommand{\ki}{k_1}              % v-link parameter in u-eqn 
\renewcommand{\L}{L}               % interval length
\renewcommand{\t}{t}               % time
\renewcommand{\u}{u}               % u-variable
\renewcommand{\v}{v}               % v-variable
\newcommand{\w}{w}                 % w-variable - complex field
\newcommand{\waven}{k}             % wavenumber in the front
\newcommand{\x}{x}                 % space variable
\newcommand{\xc}{\x_c}             % pulses "centre of mass" coordinate

\newcommand{\Fig}[1]{Fig.~\ref{fig:#1}}
\newcommand{\fig}[1]{fig.~\ref{fig:#1}}
\newcommand{\dblfigure}[3]{
  \begin{figure*}[tb!]
  \includegraphics{#1}
  \caption[]{#2}
  \label{fig:#3}
  \end{figure*}
}
\newcommand{\sglfigure}[3]{
  \begin{figure}[tb!]
  \includegraphics{#1}
  \caption[]{#2}
  \label{fig:#3}
  \end{figure}
}

\begin{document}
\title{Envelope quasi-solitons in dissipative systems with cross-diffusion}
\author{V. N. Biktashev}
\affiliation{Department of Mathematical Sciences, 
  University of Liverpool, Liverpool L69 7ZL, UK}
\author{M. A. Tsyganov}
\affiliation{
  Institute of Theoretical and Experimental Biophysics, 
  Pushchino, Moscow Region, 142290, Russia}
\date{\today}

\begin{abstract}
We consider two-component nonlinear dissipative spatially extended 
systems of reaction-cross-diffusion type.
Previously, such systems were shown to support ``quasi-soliton''
pulses, which have fixed stable structure but can reflect from
boundaries and penetrate each other. Presently we demonstrate a
different type of quasi-solitons, with a phenomenology resembling that
of the envelope solitons in Nonlinear Schroedinger equation:
spatiotemporal oscillations with a smooth envelope, with the velocity
of the oscillations different from the velocity of the envelope.
\end{abstract}
\maketitle

\section{Introduction}

Dissipative structures, \ie\ patterns in spatially extended systems
away from equilibrium have been intensively studied for many
decades. A very comprehensive review can be found in
\textcite{Cross-Hohenberg-1993}; results obtained since then would
probably require an even more extensive review.  A very popular class
of mathematical models is the reaction-diffusion systems with diagonal
diffusion matrices. There have been numerous indications that
non-diagonal elements in diffusion matrices, \ie\ cross-diffusion, can
lead to new nontrivial effects not observed in classical
reaction-diffusion systems, \eg\ \emph{quasi-solitons} in systems with
excitable reaction part \cite{QS1,QS6}. The defining
features of the quasi-solitons was their ability to penetrate each
other, which makes them akin to the true solitons in the conservative
systems.  However the question remained whether this similarity is a
reflection of common mechanisms, or is entirely superficial and
incidental. Here we report an observation which makes the similarity
even more striking.  Namely, the previously reported quasi-solitons
propagated while retaining fixed shape profile, \ie\ were constant
solutions in a co-moving frame of reference; the exceptions were the
``ageing'' quasi-solitons reported in \cite{QS6} which retained
their front and tail structures but changed their overall length.
Here we report ``envelope quasi-solitons'' (EQS), which share some phenomenology
with envelope solitons in the Nonlinear Schroedinger
Equation~\cite{Cross-Hohenberg-1993,NLS}. Namely, they have the form
of spatiotemporal oscillations (``wavelets'') with a smooth envelope, and the
velocity of the individual wavelets (the phase velocity) is
different from the velocity of the envelope (the group velocity). This
may be a serious evidence for some deep relationship between these
phenomena from dissipative and conservative realms.

Our observations are made in two two-component models, supplemented
with cross-diffusion, rather than self-diffusion terms; such terms may
appear say in 
mechanical~\cite{Cartwright-etal-1997}, 
chemical~\cite{Vanag-Epstein-2009}, 
or %
ecological~\cite[p.~11]{Murray-2003}
applications:
\begin{equation}
  \df{\u}{\t} = \f(\u,\v) + \ddf{\v}{\x}, \quad
  \df{\v}{\t} = \g(\u,\v) - \ddf{\u}{x}. \label{RXD}
\end{equation}
 We consider the FitzHugh-Nagumo (FHN) kinetics taken in the form
\begin{equation}
  \f = \u(\u-\a)(1-\u) - \ki\v, \quad
  \g = \eps \u, \label{FHN}
\end{equation}
as an archetypal excitable model,
with an arbitrarily fixed value of parameter $\ki=10$, and varied 
values of parameters $\a$ and $\eps$. 
As a specific example of a real-life system, we also consider
the Lengyel-Epstein~\cite{Lengyel-Epstein-1991}
(LE) model of chlorite-iodide-malonic acid-starch autocatalitic reaction system, 
\begin{equation}
  \f = \ale-\u - \frac{4\u\v}{1+\u^2}, \quad
  \g = \ble \left(\u - \frac{\u\v}{1+\u^2}\right), \label{LE}
\end{equation}
for fixed $\ale=6.3$ and $\ble=0.055$. 
We simulated \eq{RXD} on an interval
$\x\in[0,\L]$, $\L\le\infty$ with Neumann boundary conditions for both $u$ and
$v$~\cite{epaps}.

%%%%%%%%%%%%%%%%%%%%%%%%
\sglfigure{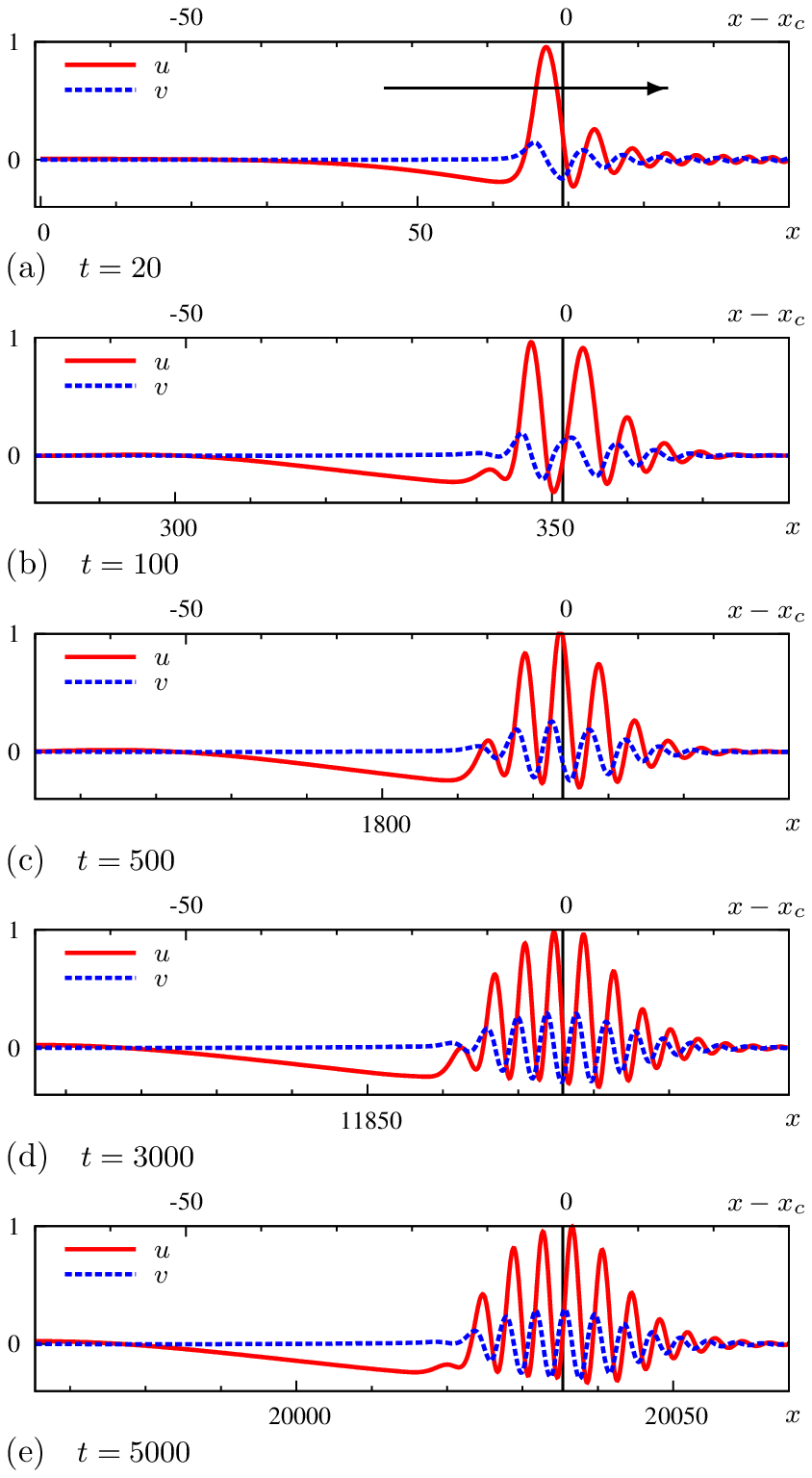}{ (color online)
  Quasi-soliton profiles at the indicated moments of time, in a co-moving
  frame of reference, upper $\x$-axes, with the reconstructed
  original spatial coordinates shown on lower $\x$-axes. 
  Parameters $\a=0.12$, $\eps=0.01$, $\L=\infty$,
  solution propagates rightwards.  (a-c) Development of a
  quasi-soliton.  (d,e) Propagation of a quasi-soliton, with unchanged
  envelope and shifting phase of high-frequency wavelets within the
  envelope. Note that wavelets in (d) and (e) are
    in antiphase: the $\v$ profile at $\x=\xc$  is near a local minimum in 
    (d) and a local maximum in (e).
}{prof}
%%%%%%%%%%%%%%%%%%%%%%%%

\Fig{prof} illustrates development and subsequent propagation of an envelope
quasi-soliton (EQS) solution in \eqtwo{RXD}{FHN}. The profiles are presented in a
co-moving frame of reference, with $x$-coordinate measured with
respect to the ``centre of mass'' $\xc$ of the quasi-soliton~\cite{epaps}.
Simulations with different initial conditions show that as long as the
initial perturbation is above a threshold, the amplitude and overall
shape of the quasi-soliton does not depend on its details. An
important feature, evident from comparing panels (d) and (e), is that
whereas the overall shape (the envelope) of the quasi-soliton and the
wavelength of the high-frequency oscillations (the wavelets) within that envelope remain
unchanged, the phase of the the wavelets relative to the envelope position
changes, so the phase velocity (of the wavelets) is different from the
group velocity (of the envelope).

%%%%%%%%%%%%%%%%%%%%%%%%
\dblfigure{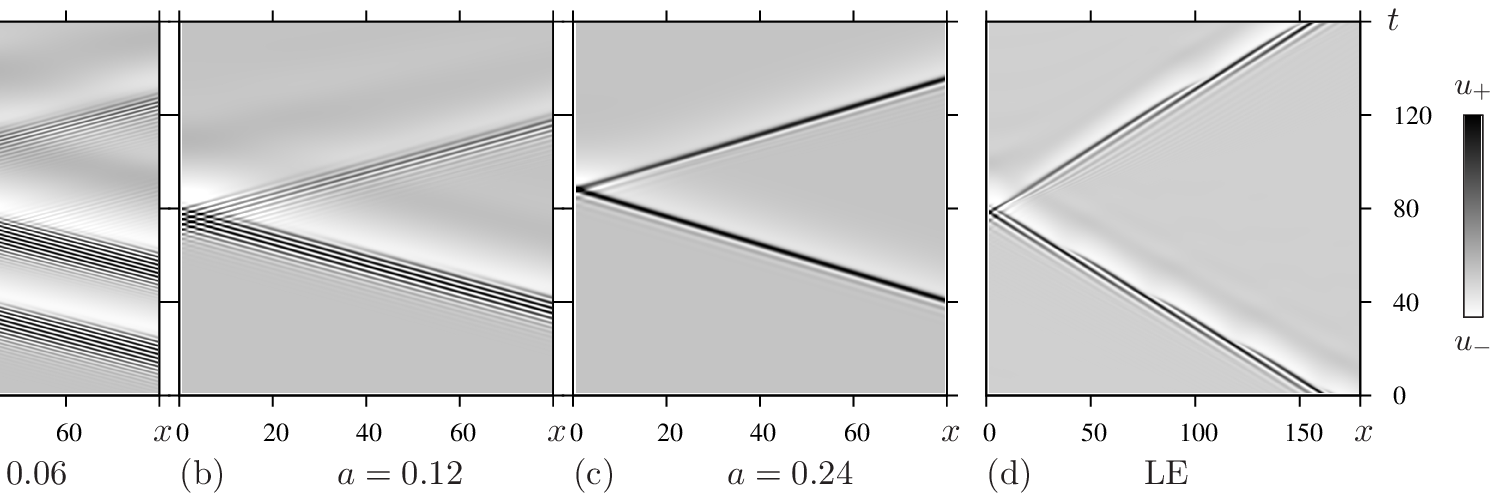}{ 
  Density plots of the quasi-solitons reflecting from a
  boundary~\cite{epaps}.
  (a-c) FHN kinetics, $u_{-}=-0.3$, $u_{+}=1$, 
  $\eps=0.01$, and $\a$ is varied as shown
  under the panels. 
  (a) A double EQS becomes a
  single EQS upon the reflection. Some
  time after that, it will grow its twin become a double quasi-soliton
  again.  
  (b) A single EQS: this is the same case
  as shown in~\fig{prof}.  
  (c) A ``classical'' quasi-soliton which retains its shape as
  it propagates.
  (d) An EQS in the LE kinetics, $\L=300$, % $\L=140$,
    $u_{-}=0.8$, $u_{+}=3.3$.
    In all panels, the origin of the $\t$-axis is shifted to an 
  arbitrarily chosen moment 
  shortly before the impact event.
}{density}
%%%%%%%%%%%%%%%%%%%%%%%%

This feature can also be seen in \fig{density}(b). The thin stripes in
the density plot represent individual wavelets, and the broader band,
consisting of these stripes, represents the envelope. The slope of the
stripes is the inverse of the phase velocity, and the slope of the
band is the inverse of the group velocity. The stripes are not
parallel to the band, because the group and the phase velocities
differ. This figure also illustrates another important phenomenon: the
reflection of the quasi-soliton from the boundary. 
 
Panels (a) and (c) in \fig{density} illustrate two different sorts of
solutions which are observed at higher and lower values of parameter
$\a$, which also reflect from the boundary. 

In panel (a), we still see individual wavelet stripes that are not
parallel to the envelope bands, but there are two bands in the
incident wave. Note that the reflected wave only has one band; however
if that reflected band is allowed to propagate for a sufficiently long
time, it will spawn its twin band behind it. This is a ``multiplying''
EQS. We do not go further into properties of these
solutions, reserving that for a future study.

In panel (c) there is only one dominant stripe and many weaker
stripes, all of which are parallel to the band. This solution has
phenomenological features similar to quasi-solitons described
previously, \eg\ in~\cite{QS1}, namely, the wave retains constant
shape as it propagates, and reflects from a boundary.

Panel (d) shows a quasi-soliton reflecting from the
  boundary, in the other model \eqtwo{RXD}{LE}. 

%%%%%%%%%%%%%%%%%%%%%%%%
\sglfigure{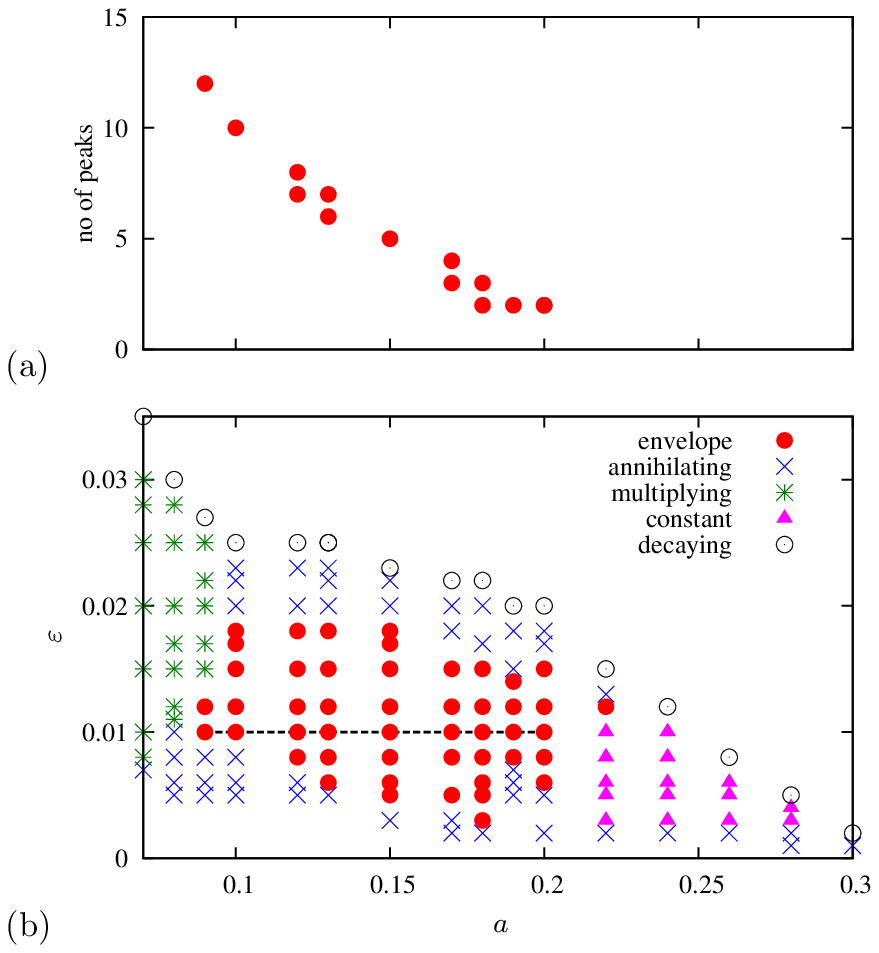}{ (color online)
  (a) The number of wavelets in an EQS as a function
    of $\a$ at fixed $\eps=0.01$. 
  (b)
  Areas of different sorts of solutions in the parametric plane
  $(\a,\eps)$. The black dashed line corresponds to the parametric
  cross-section shown in (a). 
}{areas}
%%%%%%%%%%%%%%%%%%%%%%%%

\Fig{areas} gives an overview of the parametric area of the
EQS solutions in \eqtwo{RXD}{FHN}
and its neighbours. In panel (b),
the parameter
sets at which EQS solutions like the one shown in
\fig{density}(b) have been observed, are designated by red solid
circles. This area is surrounded:
\begin{itemize}
\item 
at higher and lower values of $\eps$, by solutions which have
  similar shape to those shown in~\fig{prof} and \fig{density}(b),
  but do not reflect from boundaries (`annihilating', blue crosses);
\item
at lower values of $\a$, by multiplying EQSs,
  one of which is illustrated in~\fig{density}(a) (`multiplying',
  green stars);
\item
at higher values of $\a$, by constant shape quasi-solitons,
  such as the one shown in~\fig{density}(c) (`constant', magenta
  triangles).
\end{itemize}
The area of existence of all these solutions in the $(\a,\eps)$ parametric 
plane is bounded from above and
from the right, and beyond it our initial conditions did not produce
any stably propagating solutions (`decaying', black open circles).
Panel (a) in this figure illustrates the variability of the
  shape of EQS within their parametric domain.
Most important conclusion from~\fig{areas} is that the EQSs
are not a unique feature of a special set of parameters
but are observed in a rather broad parametric area.

The oscillatory character of the fronts of cross-diffusion waves,
described in numerical simulations~\cite{%
  QS1,%
  QS6%
} and analysed theoretically in~\cite{%
  QS1,%
  QS6,%
  Zemskov-Loskutov-2008%
}, was for waves of stationary shape.
Although the proper theory of the EQSs is beyond the scope
of this Letter, the analysis of their non-stationary front structure
is easily achieved via linearization of \eq{RXD}.
The resting states in both FHN~\eq{FHN} and
  LE~\eq{LE} kinetics are stable foci which already shows propensity to
  oscillations. Considering in more detail the 
  FHN kinetics,
for small $\u,\v$, the solution has the form
\begin{equation}
  \Mx{\u\\\v} \approx \Re{ 
    \C \evec \e^{-\decr(\x-\cg\t)} \e^{\i(\waven\x-\freq\t)} 
  },                                                        \label{fit}
\end{equation}
where
\begin{eqnarray}
  & \A(\evalt,\evals)\evec=\mx{0}, \quad \evec\ne\mx{0}, \quad \det\A=0,
                                                            \label{disp} \\
  &\A=\Mx{
    -\a-\evalt & -\ki+\evals^2 \\
    \eps-\evals^2 & -\evalt
  }, \;\evalt = \decr\cg-\i\freq,\;\evals = -\decr+\i\waven.
                                                            \nonumber
\end{eqnarray}

\sglfigure{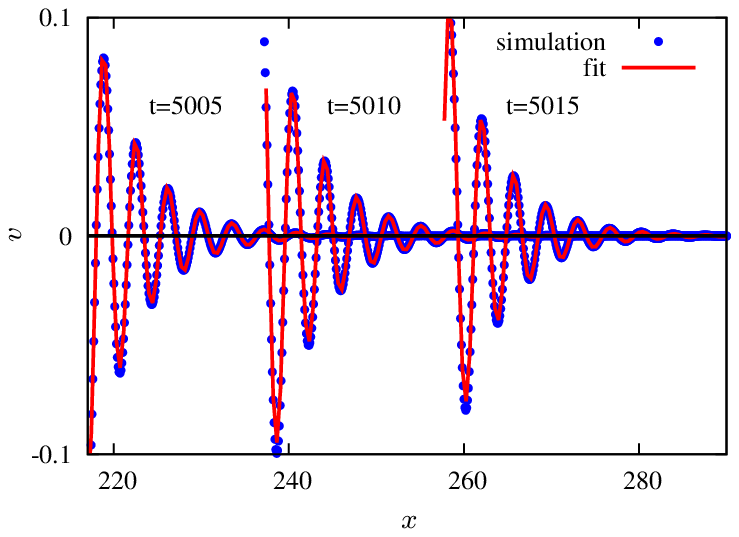}{ (color online)
  Profiles of an EQS wavefront and its fitting by~\eq{fit} at selected
  moments of time. Parameters are $\eps=0.01$, $\a=0.12$,
  $\L=\infty$. The origin of the $\x$-axis is chosen arbitrarily.
}{front}
Fitting of the $v$-component of a solution
shown on \fig{prof} to~\eq{fit} gives 
$\cg\approx 4.07698$,
$\waven\approx 1.71532$,
$\decr\approx 0.182305$ and 
$\freq\approx 6.15190$,
which satisfies~\eq{disp} to 3 s.f.~\cite{epaps} 
The quality of the fitting is illustrated in~\fig{front}. 
Note that the phase velocity
of the wavelets here is $\cp=\freq/\waven\approx3.59$,
smaller than the group velocity,
$\cg\approx4.08$, which agrees with the fact that the slope of the 
individual
stripes in
\fig{density}(b) (which is the inverse of the phase velocity $\cp$) is
steeper than the slope of the band (which is the inverse of the group
velocity $\cg$).

 The shape of the profiles in~\fig{prof} is reminiscent of
  localized states in the generalized Swift-Hohenberg equation with
  ``snakes and ladders'' bifurcation
  diagrams~\cite{Burke-Knobloch-2006}. The essential
  difference of our solutions is that they move and
  do not preserve their shape, so cannot be immediately
  studied by ODE bifurcation techniques.  

The defining features of the EQSs described above
are similar to envelope solitons of the Nonlinear Schr\"odinger
Equation. The version of this equation known as `NLS+'~\cite{NLS}, can
be written in the form
\[
\i \df{\w}{\t} + \ddf{\w}{\x} + \w|\w|^2 = 0
\]
for a complex field $\w$, which presents
a reaction-cross-diffusion system for two real fields
$\u$ and $\v$ via $w=u-\i v$ 
of the form~\eq{RXD} with 
\begin{equation}
  \f = \u(\u^2+\v^2), \quad
  \g = -\v(\u^2+\v^2).   \label{NLSsys}
\end{equation}
System~\eqtwo{RXD}{NLSsys}
has soliton solutions in the form of (fast) harmonic waves with a
unimodal ($\sech$-shaped) envelope, and the propagation velocity of
the envelope (the group velocity) is different from the propagation
velocity of the wavelets (the phase velocity). Hence one might think
of possible interpretation of the EQSs in \eqtwo{RXD}{FHN} or \eqtwo{RXD}{LE} as a
result of a non-conservative perturbation of the envelope solitons in
\eq{NLSsys}, which would select particular values of the otherwise
arbitrary amplitude and speed of the soliton and modify its
shape. This interpretation, however, does not seem to work, and our
attempts to connect the solutions in \eqtwo{RXD}{NLSsys} and \eqtwo{RXD}{FHN}  via a
one-parametric family of systems have been unsuccessful,
as the EQS solutions disappeared during
  parameter continuation. The apparent
reason is that the sense of rotation of solutions of \eq{NLSsys} in
the $(\u,\v)$ is clockwise whereas in \eq{FHN} and \eq{LE} it is
counterclockwise, and the variant of \eq{NLSsys} with counterclockwise
rotation, the `NLS-' equation, does not have envelope soliton
solutions.

Another comparison can be made with ``wave packets'' reported by Vanag
and Epstein in microemulsion BZ reaction, and corresponding
mathematical models, associated with finite-wavelength instability of
an equilibrium in a reaction-diffusion system with unequal
self-diffusion coefficients~\cite{%
 Vanag-Epstein-2004%
}. They considered two distinct types of solutions: small- and
large-amplitude wave packets (SAWP and LAWP), both capable of
reflection from boundaries. SAWP are observed in the nearly-linear regime,
they have the phase speed (of the wavelets) different from the group speed
(of the envelope, or the packet). However, being near to a linear
instability and having no stabilizing effect of the dispersion as in
NLS+, the packets slowly grow both in amplitude and in width, \ie\
they are not quasi-solitons. The LAWP, on the contrary, have fixed
amplitude and width, but their phase and group velocities coincide, so
they retain constant shape. They are therefore phenomenologically
similar to the quasi-solitons reported in excitable systems with
cross-diffusion~\cite{QS1}. Note that adiabatic elimination of a fast
component in a reaction-diffusion system with very different
self-diffusion coefficients is one of the ways in which
cross-diffusion terms may appear~\cite{Kuznetsov-etal-1994,Vanag-Epstein-2009}, so this
analogy deserves further investigation.

To conclude, the solutions we have reported resemble NLS+ envelope
solitons by their morphology and by their ability to reflect from
boundaries, however they are different in that amplitudes and speeds of
NLS solitons depend on initial conditions, while in \eq{RXD} they
are fixed by parameters of the models. The reported solution are
similar to quasi-solitons reported earlier in that they share the
fixed amplitude and reflection properties, but different in that they
do not preserve constant shape as their phase
velocities are different from their group velocities. Hence we believe this is a new
nonlinear phenomenon not seen before. The mechanisms behind the key
properties of this new type of solutions require further
investigation, however it is already clear that this is not simply a
non-conservative perturbation of NLS.

\textbf{Acknowledgments}
The study was supported in part 
by a grant from the Research Centre for Mathematics and Modelling,
University of Liverpool (UK).

%%%%%%%%%%%%%%%%%%%%
%\bibliography{eqs} 

\end{document}